\documentclass[doublecol]{epl2}

\usepackage{amsmath}
\title{Detection of interaction-induced nonlocal effects using 
perfectly transmitting nanostructures}
\author{Dietmar Weinmann\inst{1} \and Rodolfo A. Jalabert\inst{1} \and
Axel Freyn\inst{2} \and Gert-Ludwig Ingold\inst{3} \and Jean-Louis 
Pichard\inst{2}}
\institute{
  \inst{1} Institut de Physique et Chimie des Mat{\'e}riaux de Strasbourg,
  UMR 7504 (CNRS-ULP) - 23 rue du Loess, BP 43, F-67034 Strasbourg Cedex 2,
  France\\
  \inst{2} Service de Physique de l'{\'E}tat Condens{\'e} (CNRS URA 2464),
  DSM/IRAMIS/SPEC, CEA Saclay - F-91191 Gif sur Yvette Cedex, France\\
  \inst{3} Institut f{\"u}r Physik, Universit{\"a}t Augsburg - D-86135
  Augsburg, Germany
}
\shortauthor{D. Weinmann \etal}


\abstract{%
We consider one-dimensional transport through an interacting region in 
series with a point-like one-body scatterer. When the conductance of 
the interacting region is perfect, independently of the interaction 
strength, a nonlocal interaction effect yields a total conductance of 
the composed system that depends on the interaction strength and is 
lower than the transmission of the one-body scatterer. This qualitative 
nonlocal effect allows to probe the dressing cloud of an interacting 
system by ideal noninteracting leads. The conductance correction 
increases with the 
strength of the interaction and the reflection of the one-body scatterer 
(attaining relative changes $>50\%$), and decreases with the distance 
between the interacting region and the one-body scatterer. Scaling laws 
are obtained and possible experimental realizations are suggested.
}

\begin{document}

\maketitle

\section{Introduction}

Viewing quantum transport as a scattering problem is at the heart of 
Landauer's approach to the conductance of mesoscopic systems \cite{datta}. 
Working at zero temperature and ignoring electron-electron interactions we 
solely have to consider the elastic scattering of electrons at an 
energy that is given by the Fermi level of the electrodes, arising from 
one-body potentials in the system. 
Including interactions through Landau quasiparticles does not 
modify appreciably this situation. 
Such an effective one-body description allows to understand a wealth of 
phenomena, ranging from residual resistivity to interference effects.

Interaction effects become prominent in small and weakly connected quantum 
dots displaying Coulomb blockade oscillations \cite{datta}.
Describing the charging effects through capacitances one stays at the level 
of a simplified local mean-field approach, and the view of quantum transport 
as the scattering of (quasi-)particles is still applicable. 
Exploring yet smaller systems within the hypothesis of an effective one-body 
scattering we encounter cases in which the effective transmission can no 
longer be obtained from such a simple approach ignoring many-body exchange 
and correlation effects. 
Such many-body signatures have been theoretically demonstrated for the 0.7 
anomaly of quantum point contacts \cite{sushkov01}, in the length-dependent
oscillations of the conductance through an atomic chain \cite{molina04b}, and
in the interaction-induced increase of the conductance through a strongly 
disordered quantum wire \cite{molina03}.    
In realistic systems it is difficult to control the effective interaction 
strength independently of other parameters. Thus, the clear-cut observation 
of many-body signatures on the measured transmissions has remained elusive. 

For systems containing a small region, that we refer to as \textit{nanosystem} 
and in which interactions are important, the situation is even more 
complicated since one can pose the fundamental question under which 
circumstances the one-body scattering approach yielding an effective 
transmission is valid.  
Kondo physics in the transport through ultra-small quantum dots 
\cite{goldhaber98,pustilnik01} provides an example where electronic 
correlations are necessary for the interpretation of the data and where 
one-body concepts cannot be used inside the spin screening cloud 
\cite{hewson_book,bergmann}.

In generic nanosystems, the effective one-body approaches are challenged by 
the nonlocal effects arising from interactions which can be tested by 
approaching an external scatterer. The nonlocality can be explained already 
at the Hartree-Fock (HF) level 
\cite{asada-freyn-pichard,freyn07a_1,freyn07a_2}, since the Hartree and 
Fock corrections are given by nonlocal coupled integral equations. For 
instance, the effect of an external scatterer upon the Hartree corrections 
results from the Friedel oscillations of the electron density that the 
external scatterer induces inside the nanosystem. 
The Fock corrections are characterized by similar oscillations 
\cite{asada-freyn-pichard}. 
Using a model where particle-hole symmetry yields a uniform density, 
the nonlocal effect arises from the exchange. 
At zero temperature, introducing an external scatterer at a distance 
$L_\mathrm{C}$ of the nanosystem yields an effect which decays as the Friedel 
oscillations which cause it, i.e.\ $\propto 1/L_\mathrm{C}$ in one dimension. 
At a finite temperature $T$, this effect is exponentially suppressed 
\cite{asada-freyn-pichard}, if $L_\mathrm{C}$ exceeds the thermal length 
$L_\mathrm{T}$ describing the scale over which the electrons at the Fermi 
surface propagate during a time $\hbar/k_BT$. 
This means that all the external scatterers located in a region of size 
$L_\mathrm{T}$ can modify the transmission of the interacting nanosystem. 

Only the \textit{dressed nanosystem}, consisting of the interacting region and 
its local environment, behaves as an effective one-body scatterer. 
The fundamental question posed above translates into the question about the 
nature and the extension of the associated cloud dressing the nanosystem.
A way of testing such a cloud is to study the conductance through two 
nanosystems connected in series by a short noninteracting lead, 
and to detect the deviations from the prediction based on the combination 
of effective one-body scatterers \cite{molina05}. 
This is in principle an experimentally observable effect, 
although it seems difficult to get simultaneous access to both, the 
conductance of the individual and of the combined system, and/or to change 
the separation between the interacting regions.

Within this line of investigations it seems more promising to replace one 
of the nanosystems by a tunable one-body scatterer.
If the one-body scatterer is influenced by an attached Aharonov-Bohm ring,
a HF treatment indicates that a nonlocal interaction effect can lead
to a significant dependence of the nanosystem transmission on the magnetic 
flux piercing the ring \cite{freyn07a_1,freyn07a_2}.
If the one-body scatterer is a scanning gate microscope 
acting on a two-dimensional electron gas in 
the proximity of a quantum point contact \cite{westervelt_1,westervelt_2}, 
its effect upon the resulting 
conductance carries the signature of the electron-electron interactions 
inside the constriction \cite{freyn08}. In these two cases the interaction 
determines both, the effective transmission of the nanosystem as well as 
the interaction-induced nonlocal correction to the total conductance. 
However, if the conductance of the nanosystem is independent of 
the interaction strength, the only source of interaction dependence of 
the total conductance can be the nonlocal interaction effect. This is a 
striking situation because an interaction-dependent conductance implies the 
nonapplicability of the standard composition law and therefore immediately 
demonstrates the presence of nonlocal effects.

In this work we consider precisely this situation by setting up a 
one-dimensional model with parameters chosen such that the 
transmission through the nanosystem is perfect for all values of the 
interaction strength \cite{molina03,molina04b}, while the one-body point-like 
scatterer is introduced as an electrostatic perturbation.

\section{Nanosystem with perfect conductance}

\begin{figure}
\centerline{\includegraphics[width=0.9\columnwidth]{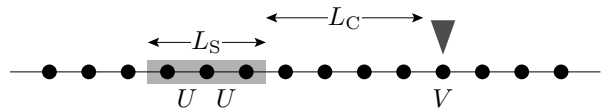}}
\caption{\label{fig:intimp-sketch}One-dimensional setup with an interacting 
region (nanosystem, grey) of length $L_\mathrm{S}$ and a local one-body 
scattering potential $V$ at a distance $L_\mathrm{C}$.}
\end{figure}

We consider spinless fermions in a one-dimensional chain with an interacting 
region of length $L_\mathrm{S}$, separated by a lead of length $L_\mathrm{C}$ 
from a point-like scatterer (see Fig.~\ref{fig:intimp-sketch}). 
The corresponding Hamiltonian reads
\begin{equation}\label{hamiltonian}
H = H_\mathrm{kin} + H_\mathrm{int} + H_\mathrm{1bs} \, , 
\end{equation}
where 
\begin{equation}
H_\mathrm{kin} = - \sum_{i=-\infty}^{\infty} 
\left( c_i^\dagger c_{i+1}^{\phantom{\dagger}} + \mathrm{h.c.} \right)
\end{equation}
is the kinetic energy part. 
Here, $c_{i}$ annihilates a fermion on site $i$, and we have fixed the energy
scale by setting the hopping amplitude equal to unity.
The nearest-neighbour interaction on sites 1 to $L_\mathrm{S}$ is described by 
\begin{equation}
H_\mathrm{int} = U \sum_{i=1}^{L_\mathrm{S}-1} \left(n_{i} - 1/2\right) 
\left(n_{i+1}-1/2\right) 
\end{equation}
with the local density operators $n_i=c_i^\dagger c_i^{\phantom{\dagger}}$.

At half filling, an odd number of sites $L_\mathrm{S}$ ensures a perfect 
effective transmission, i.e.~a dimensionless conductance $g=G/(e^2/h)=1$, for 
the nanosystem, independent of the interaction strength 
\cite{molina03,molina04a,molina04b}. 
We choose half filling and $L_\mathrm{S}=3$ in order to keep the size of the 
total system as small as possible.

The one-body on-site scattering potential of height $V$ that is separated 
from the interacting region by $L_\mathrm{C}$ sites is represented by
\begin{equation}\label{vham}
H_\mathrm{1bs} = V\,n_{L_\mathrm{S}+L_\mathrm{C}+1}\, .
\end{equation}

It is straightforward to calculate the transmission probability and thus the 
conductance $g_\mathrm{1bs}$ of an on-site potential of strength $V$ in a 
clean chain. 
At energy $E=0$ corresponding to the Fermi energy at half filling, 
one gets 
\begin{equation}
g_\mathrm{1bs}=\frac{4}{4+V^2}\, .
\end{equation}
For two one-body scatterers in series, where the first one is
characterized by perfect transmission, the total transmission is simply given 
by the transmission of the second scatterer. 
In our case, one would therefore naively expect a total conductance 
$g=g_\mathrm{1bs}$. 
However, the perfect transmission of the interacting region is an effective 
transmission describing the interacting region including long attached leads.
The presence of a one-body scatterer in the vicinity of the 
nanosystem affects its transmission. We will show that this leads 
to pronounced deviations of the total conductance from $g_\mathrm{1bs}$.

\begin{figure}
\centerline{\includegraphics[width=0.9\columnwidth]{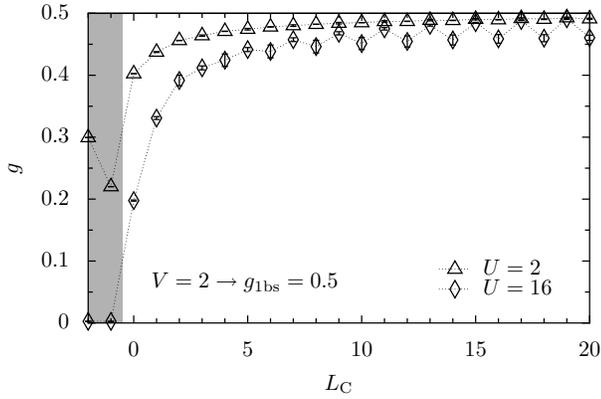}}
\caption{\label{fig:g-LcV2}Conductance of the combined system sketched 
in Fig.~\ref{fig:intimp-sketch} as a function of $L_{\rm C}$ for a 
scatterer with $V=2$ for $U=2$ (triangles) and $U=16$ (diamonds). The grey area
indicates that the scatterer is placed in the interacting region.}
\end{figure}

\begin{figure}
\centerline{\includegraphics[width=0.9\columnwidth]{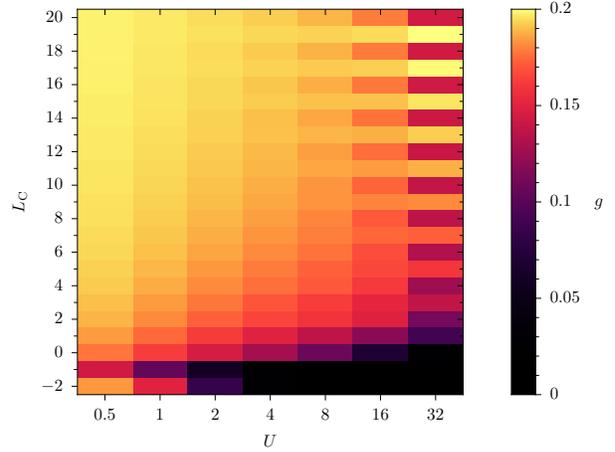}}
\caption{\label{fig:g-ULcV4}Conductance $g$ as a function of $L_{\rm C}$ and 
$U$ for a scatterer with strength $V=4$ corresponding to $g_\mathrm{1bs}=0.2$.}
\end{figure}

The embedding method 
\cite{favand98,sushkov01,molina03,meden03,rejec03,molina04a} 
allows to calculate the zero-temperature linear conductance through the 
system composed of the interacting part and the one-body scatterer in series. 
Within this method the conductance is extracted from the charge stiffness 
of a ring composed by the system and a long noninteracting lead in the 
limit of infinite lead length. 
We use the DMRG algorithm \cite{DMRG_book,schmitteckert_thesis} to determine 
the stiffness $D$ for different ring sizes $L$ up to 120 sites and 
extrapolate the results to infinite system size using fits of second-order 
polynomials to the numerical data for $\log D$ as a function of $1/L$. 
An estimate for the precision of the resulting extrapolated value is given 
by the difference of the result as compared to the one obtained from a 
linear extrapolation of $\log D (1/L)$. The resulting precision of the 
extracted conductance is displayed by the error bars in some of the figures. 
Using the DMRG algorithm allows to obtain exact results for 
finite system sizes in contrast to the HF approximation,
which fails when $U$ or $L_\mathrm{S}$ become large and induce nonnegligible 
correlation effects \cite{asada-freyn-pichard}.

\begin{figure}
\centerline{\includegraphics[width=0.9\columnwidth]{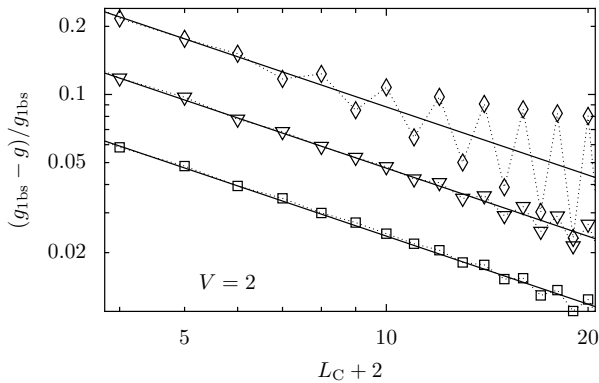}}
\caption{\label{fig:g-LcV2fit}Log-log plot of the relative conductance 
correction as a function of $L_{\rm C}+2$ at $V=2$, 
for $U=1$ (squares), $U=4$ (triangles), and $U=16$ (diamonds). 
The full lines are fits of \eqref{eq:lcscaling} while the dotted lines
serve to guide the eye.}
\end{figure}
   
The dependence of the conductance $g$ on the length $L_\mathrm{C}$  
for an on-site potential $V=2$ is shown in Fig.~\ref{fig:g-LcV2} for two 
different values of the interaction strength. 
In this case, the conductance of the one-body scatterer is 
$g_\mathrm{1bs}=0.5$. 
Pronounced deviations from that value appear at small separations between 
the scatterer and the nanosystem. 
Those deviations are negative and increase with decreasing $L_\mathrm{C}$. 
While this seems reminiscent of the increase of the deviations observed 
for even values of $L_\mathrm{C}$ when two nanosystems in series are 
considered \cite{molina05}, there is nevertheless an important difference. 
In the present case, deviations from $g_\mathrm{1bs}$ appear for all values 
of $L_\mathrm{C}$. 
For not too strong interaction strength $U=2$, the deviations increase 
to 20\% of $g_\mathrm{1bs}$ as $L_\mathrm{C}$ decreases down to zero.
For $U=16$, the effect is stronger (60\% of $g_\mathrm{1bs}$) and 
even-odd oscillations as a function of $L_\mathrm{C}$ appear for large 
values of $L_\mathrm{C}$. The negative values of $L_\mathrm{C}$ indicated 
in grey correspond to the case 
where the potential scatterer is located inside the nanosystem. 
The deviation from the noninteracting conductance is particularly strong 
for $L_\mathrm{C}=-1$, when the one-body scatterer acts at the edge of the 
three interacting sites.

The behaviour of the total conductance $g$ as a function of $U$ and 
$L_\mathrm{C}$ is shown in Fig.~\ref{fig:g-ULcV4} for $V=4$, 
that corresponds to a conductance of $g_\mathrm{1bs}=0.2$ for the one-body 
scatterer alone. The total conductance is equal to $g_\mathrm{1bs}$ in the 
noninteracting case ($U=0$) and starts to decrease with increasing 
$U$. In addition, it can be seen that the oscillations of $g$ with
$L_\mathrm{C}$ appear already for moderate interaction strength. 
These oscillations and the deviations of the total conductance $g$ from 
$g_\mathrm{1bs}$ become stronger as $U$ increases. As a consequence, 
$g$ assumes large values approaching $g_\mathrm{1bs}$ for large $U$ when 
$L_\mathrm{C}$ is odd.

The $U$-dependence of the deviations of $g$ from $g_\mathrm{1bs}$ 
contrasts with the situation for two interacting regions in series 
\cite{molina05} where the deviations from the noninteracting composition 
law of scatterers reach a maximum around $U=2$ and decrease for stronger 
values of $U$. 
The qualitative difference in the behaviour arises from the fact that 
in the case considered in Ref.~\cite{molina05}, for even $L_\mathrm{C}$
the conductance of the interacting nanosystems and the total conductance 
decrease with increasing interaction strength. 
This results in a decrease of the nonlocal corrections, which are 
further reduced by the effective decoupling of the nanosystem from 
the noninteracting leads occurring in the limit of strong interactions 
\cite{vasseur06}.  

Apart from the oscillations which are particularly noticeable at strong 
interaction, the decrease of the nonlocal correction with 
$L_\mathrm{C}$ is quite well described by the scaling 
\begin{equation}\label{eq:lcscaling}
\frac{g_\mathrm{1bs}-g}{g_\mathrm{1bs}}=\frac{A(U,V)}{L_\mathrm{C}+2} . 
\end{equation} 
This is shown in Fig.~\ref{fig:g-LcV2fit} and confirms the expected scaling 
due to Friedel oscillations caused by the one-body scatterer influencing 
the nanosystem whose centre is at a distance  
$L_\mathrm{C}+2$ from the perturbing tip. 

\section{Potential scatterer \textit{versus} weak link}
 
The potential scatterer studied so far breaks particle-hole symmetry, and
therefore disturbs the uniform electron density at half filling in the 
perfectly transmitting nanosystem. It is then important to study whether the
nonlocal effect changes when the one-body scatterer does not break
particle-hole symmetry. This can be achieved by using a weak link as a
one-body scatterer, where the term
$H_\mathrm{1bs}$ in the Hamiltonian \eqref{hamiltonian} is set to 
\begin{equation}\label{wlham}
H_\mathrm{1bs}=(1-t_\mathrm{wl})\left(c_{L_\mathrm{S}+L_\mathrm{C}+1}^{\dagger}
c_{L_\mathrm{S}+L_\mathrm{C}}^{\phantom{\dagger}}
+\mathrm{h.c.}\right)\, ,
\end{equation}
replacing the hopping matrix element for the link between the sites 
$L_\mathrm{S}+L_\mathrm{C}$ and $L_\mathrm{S}+L_\mathrm{C}+1$ 
by $t_\mathrm{wl}$. For the conductance of the weak link alone at the Fermi
energy one obtains 
\begin{equation}
g_\mathrm{1bs}=\frac{4}{\left(t_\mathrm{wl}+1/t_\mathrm{wl}\right)^2}\, .
\end{equation}  

\begin{figure}
\centerline{\includegraphics[width=0.9\columnwidth]{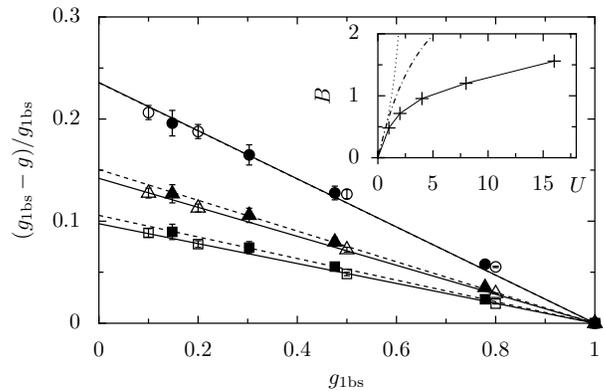}}
\caption{\label{fig:gr-g0Lc3}Relative change of the conductance 
as a function of the conductance of the one-body scatterer arising from  
an on-site potential (empty symbols; $V=1$, 2, 4, and 6)
and a weak link (filled symbols; $t_\mathrm{wl}=0.2$, 0.3, 0.4, and 0.6), 
at $L_\mathrm{C}=3$. Squares, triangles and circles represent data for
$U=1$, 2, and 8, respectively. Linear fits of the form \eqref{eq:linfit} 
are shown for an on-site potential and a weak link as solid and dashed
lines, respectively. Inset: Parameter $B$ of 
the fit \eqref{eq:linfit} for the case of a potential scatterer 
as a function of $U$ for $L_\mathrm{C}=1$. The results for $L_\mathrm{C}=3$
and 5 as well as for a weak link collapse on the same line with deviations
below 3.2\%. HF results at $L_\mathrm{C}=1$ for the slope at 
$g_\mathrm{1bs}=1$ are shown for a potential scatterer (dotted line) and 
a weak link (dashed-dotted line).}
\end{figure}
The results for the conductance of the combined system are very similar to 
the situation of a potential scatterer in series with an interacting region. 
In Fig.~\ref{fig:gr-g0Lc3} we compare the relative change of the 
conductance due to the interactions as a function of the uncorrelated 
conductance $g_\mathrm{1bs}$ of the one-body scatterer 
alone, for $L_\mathrm{C}=3$. 
We plot the dependence of the relative change  
$(g_\mathrm{1bs}-g)/g_\mathrm{1bs}$ of the total conductance on the 
conductance $g_\mathrm{1bs}$. 
The full symbols are data points obtained for a potential scatterer 
described by \eqref{vham} while the open symbols stand for a system where 
the one-body scatterer is modeled as a weak link \eqref{wlham}. 
The solid and dashed lines are the corresponding linear fits.
The two different ways of modeling the 
one-body scatterer yield nonlocal effects that are very close, 
demonstrating that the nonlocal correlation effect scales with 
the conductance $g_\mathrm{1bs}$, independent 
of the nature of the one-body scatterer.
As expected, the relative change of the conductance due to the interaction 
effect increases monotonically with decreasing $g_\mathrm{1bs}$. 

Already for $L_\mathrm{C}=3$ the nonlocal effect can amount to a 
conductance change of more than 20\%. This value increases beyond 
60\% for $L_\mathrm{C}=0$, when the on-site potential is applied on 
the first noninteracting site. 

For not too strong interaction, the dependence follows approximately the 
linear relationship
\begin{equation}\label{eq:linfit}
\frac{g_\mathrm{1bs}-g}{g_\mathrm{1bs}}=
B(U)\frac{1-g_\mathrm{1bs}}{L_\mathrm{C}+2}\, ,
\end{equation}
yielding $A(U,g_\mathrm{1bs})=B(U)(1-g_\mathrm{1bs})$ for the parameter 
$A$ in \eqref{eq:lcscaling}.
For the whole range of explored parameters, the $U$-dependence of $B$, 
shown in the inset of Fig.~\ref{fig:gr-g0Lc3}, is monotonically increasing. 
The collapse of the data on a universal 
curve confirms the scaling \eqref{eq:lcscaling}. 

The scaling law \eqref{eq:linfit} represents evidence for an intrinsic 
property of the cloud dressing the nanosystem, namely that an external 
scatterer placed in the proximity of the perfectly transmitting 
nanosystem yields a universal renormalized conductance $g$ given by 
\eqref{eq:linfit}.

A comparison of the quasi-exact results obtained using DMRG 
to HF results (grey lines) is shown in the inset of Fig.~\ref{fig:gr-g0Lc3}.
As it turns out that the linearity of the scaling \eqref{eq:linfit} is
not satisfied by the HF results, the parameter $B$ is deduced from the
slope at $g_\mathrm{1bs}=1$. The linear scaling of the nonlocal conductance 
correction with $g_\mathrm{1bs}$, independent of the nature of the scatterer, 
persists in the quasi-exact results beyond $U\simeq 10$, while HF yields 
different results for different scatterers having the same $g_\mathrm{1bs}$. 
Since the HF results exhibit the universality of the exact results 
only at rather weak interaction, the cloud dressing the nanosystem 
carries the signature of electronic correlations. 

\section{Nanosystem with interaction-dependent conductance}

\begin{figure}
\centerline{\includegraphics[width=0.9\columnwidth]{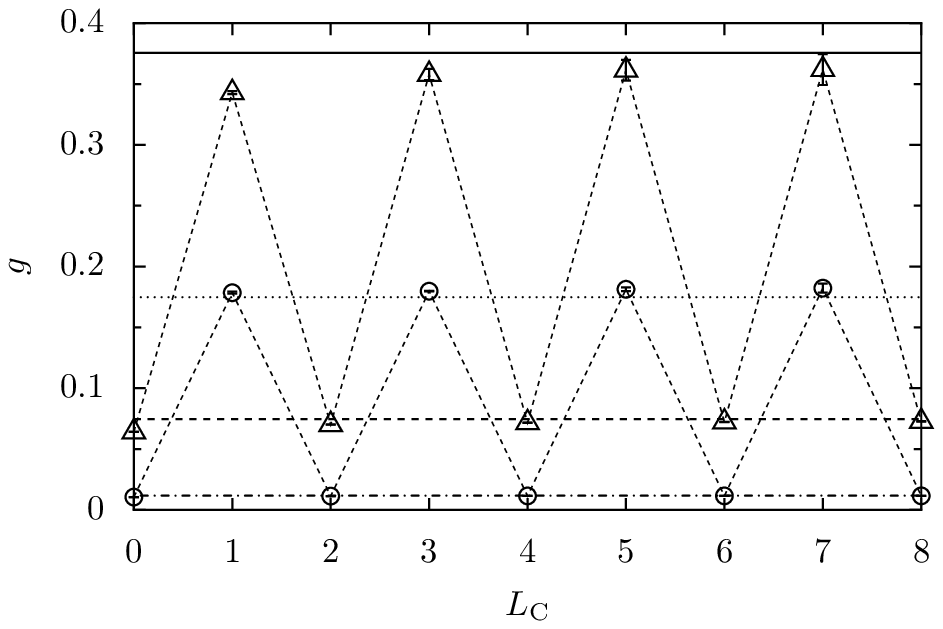}}
\caption{\label{fig:Ls2}Conductance $g$ as a function of $L_\mathrm{C}$
for an interacting region of length $L_\mathrm{S}=2$ and a potential 
scatterer with $V=4$. Triangles and circles are for $U=2$ and 8, respectively. 
The horizontal lines are the predictions using the noninteracting 
composition law of scatterers.} 
\end{figure}

The results presented above show that the nonlocal effects in the conductance 
appear in a particularly spectacular fashion when the nanosystem has perfect 
transmission. 
When the transmission through the nanosystem is interaction-dependent, 
the signature of nonlocal interaction effects has to be extracted from the 
difference between the total conductance and the prediction resulting from 
the composition of the effective interaction-dependent scatterer corresponding
to the nanosystem with the one-body scatterer.
Fig.~\ref{fig:Ls2} presents the case where the electrons interact only 
inside a nanosystem of length $L_\mathrm{S}=2$, such that the effective 
transmission of the nanosystem depends on $U$ \cite{molina03,molina04b}.    
The total conductance exhibits strong even-odd oscillations as a function of 
$L_\mathrm{C}$ that are most pronounced when the values of the transmissions 
of the one-body scatterer and the nanosystem are close, such 
that the noninteracting composition law predicts Fabry-P{\'e}rot-like 
oscillations. The deviations of the total conductance from the 
noninteracting composition law represented by the horizontal lines are 
much smaller than for the case of a perfectly transmitting interacting 
nanosystem. This example shows that choosing a nanosystem with perfect 
conductance allows for a qualitative effect of nonlocal interactions, 
unlike the merely quantitative corrections in the general case.

\section{Discussion}

We have demonstrated that a nanosystem connected in series with a one-body 
scatterer constitutes an ideal configuration to identify the nonlocality 
of the transmission in the presence 
of interaction and to detect how it is dressed by the attached leads. 
In particular, when the transmission through the nanosystem is perfect, the 
nonlocal interaction effects can be unambiguously identified because they 
result in dependencies of the conductance of the total system on the 
interaction strength and on the position of the one-body scatterer. 

The experimental confirmation of this striking effect necessitates a 
perfectly transmitting nanostructure together with the ability to control the 
interaction strength and/or the distance between the nanostructure and 
the one-body scatterer. 
Silicon quantum wires with nanosize MOSFETs allow to define regions with 
strong local enhancement of the effective electron-electron interaction 
\cite{sanquer}.
However, considerable disorder is present in the case of Ref.\
\cite{sanquer}, and
our predictions are not directly testable at present in this kind of 
structures.
A clean quantum wire with a single occupied transverse
channel represents a
possible realization of a one-dimensional model that has 
been achieved using cleaved edge overgrowth \cite{yacobi} or by local 
oxidation \cite{kvon} techniques in GaAs-GaAlAs heterostructures.
Well defined conductance plateaus as a function of the gate voltage are 
obtained in Ref.~\cite{yacobi}, and the single-mode regime is reached. 
Gating a part of the wire will allow to vary locally the electron 
density leading to an increased importance of the interactions close to 
the gate and thus defining our nanosystem. 
However, the screening induced by the gate might weaken the 
increase of the effective electron-electron interaction arising from the 
low local density \cite{shklovskii}.    
If the gate is not too close to the quantum wire, the importance of the 
interactions increases only gradually along the wire as we 
approach the nanosystem. 
Therefore the conductance of the nanosystem can be expected to be perfect, 
independent of the precise value of the interaction strength 
\cite{maslov,safi,molina03}.
A nearby scanning gate microscope (SGM) would correspond 
to the one-body scatterer of our model.

A dependence of the total conductance of a perfectly transmitting nanosystem 
on the distance between the nanosystem and the tip of the SGM will be a clear 
consequence of the nonlocal interaction effect. In a quantum point contact 
close to pinch-off, strong values of the interaction strength up 
to $U\approx 2\pi$ result from estimations of the screened on-site Coulomb 
interaction using a two-dimensional setup \cite{richter}. 
This estimation, which is conservative for the one-dimensional case of 
interest falls in the range of interaction strengths where we observe large 
oscillations of the total conductance with the position of the tip (see 
Figs.~\ref{fig:g-LcV2}, \ref{fig:g-ULcV4}). 
For these values of the interaction strength, considering only mean-field 
and exchange effects becomes unreliable in our one-dimensional models, 
and more exact methods like DMRG are needed. 

A complementary test in clean quantum wires would be to vary the gate 
voltage from close-to-open to close-to-pinch-off, thus changing the 
electron density and the strength of the effective interaction while 
remaining in a single-channel situation. 
In that case we expect to observe a gate-voltage dependence of the total 
conductance, that would be absent if the tip were removed.     

The impressive advances of SGM allow to envision other 
tests of our model. 
Recently this experimental technique has been applied to more complicated 
setups, like the proximity of a quantum point contact \cite{aoki}, the 
imaging of a one-electron quantum dot in a nanowire \cite{westervelt08},
and Aharonov-Bohm rings \cite{sellier}.
In the latter experiments the imaging of wave functions inside open quantum 
rings was achieved.
Numerical calculations neglecting interactions yield patterns resembling the 
observed ones. 
Such a correspondence might be explained by the fact that the rings are 
relatively large and contain many transverse channels, such that interaction 
effects might not be very important. 
Going to smaller structures and eventually to the single-channel 
configuration will enhance the role of interactions, thus yielding a 
nanosystem in the sense of our interacting model region. 
We have seen that positioning a one-body scatterer inside the nanosystem 
(negative $L_\mathrm{C}$ in Figs.~\ref{fig:g-LcV2} and \ref{fig:g-ULcV4}) is 
also a way of exploring electronic interactions.
The conductance obtained when the tip is close or inside the nanosystem 
can be very different from the conductance resulting only from the 
backscattering by the tip.

In summary, we have demonstrated the importance of nonlocal interaction 
effects in quantum transport through nanostructures.
These effects are particularly striking when the dressing cloud of a 
nanosystem with perfect transmission is perturbed by a one-body scatterer.
We have suggested experimental setups in which the predicted effects can
be detected.

\acknowledgments{We thank P.~Schmitteckert for his DMRG code and useful 
discussions. Financial support has been provided by the European
Union through the MCRTN program (contract MCRTN-CT-2003-504574).}

\end{document}